\documentclass[11pt]{article}
\usepackage{amssymb,amsmath,amsfonts}
\usepackage{graphicx}
\usepackage{graphics}
\usepackage{eepic,epsfig}

\begin{document}

\title{COSMOLOGICAL EVOLUTION WITH NEGATIVE\\
ENERGY DENSITIES}
\author{A. A. SAHARIAN$^{1,2}$\thanks{%
E-mail: saharian@ysu.am}, R. M. AVAGYAN$^{1,2}$, E. R. BEZERRA DE MELLO$^3$,
\and V. KH. KOTANJYAN$^{1,2}$, T. A. PETROSYAN$^{1,2}$, H. G. BABUJYAN$^1$
\vspace{0.3cm} \\
\textit{$^1$Department of Physics, Yerevan State University, }\\
\textit{1 Alex Manogian Street, 0025 Yerevan, Armenia }\vspace{0.3cm}\\
\textit{$^2$Institute of Applied Problems of Physics NAS RA, }\\
\textit{25 Hr. Nersessian Street, 0014 Yerevan, Armenia }\vspace{0.3cm}\\
\textit{$^3$Departamento de F\'{\i}sica, Universidade Federal da Para\'{\i}%
ba,} \\
\textit{58059-900, Caixa Postal 5008, Jo\~{a}o Pessoa, PB, Brazil}}
\maketitle

\begin{abstract}
For general number of spatial dimensions we investigate the cosmological
dynamics driven by a cosmological constant and by a source with barotropic
equation of state. It is assumed that for both those sources the energy
density can be either positive or negative. Exact solutions of the
cosmological equations are provided for flat models. For models with curved
space and with zero cosmological constant the general solutions are
expressed in terms of the hypergeometric function. The qualitative evolution
is described for all values of the equation of state parameter. We specify
the values of that parameter and the combinations of the signs for the
cosmological constant and matter energy density for which the cosmological
dynamics is nonsingular. An example is considered with positive cosmological
constant and negative matter energy density induced by the polarization of
the hyperbolic vacuum.
\end{abstract}

\bigskip


\bigskip

\section{Introduction}

The investigation of cosmological dynamics is carried out mainly within the
framework of homogeneous and isotropic models described by
Friedmann-Robertson-Walker (FRW) line element. In particular, the models
containing a positive cosmological constant in addition to the matter and
radiation sources of the expansion have been actively studied. This
theoretical activity is motivated by the observational evidence \cite%
{Ries98,Perl99} for accelerated expansion of the universe in recent epoch
driven by a source (dark energy) with properties close to a positive
cosmological constant. The cosmological model with a positive cosmological
constant and cold dark matter (CDM) in addition to the usual matter ($%
\Lambda $CDM model) is in good agreement with observational data on the
large scale structure and dynamics of the universe. Recently a problem
appeared that is related to the value of the Hubble parameter $H_{0}$ at
present determined by two different ways. The first one is based on direct
low redshift observations \cite{Free19}-\cite{Ries21} and gives the value $%
H_{0}\approx 73\,\mathrm{km/s/Mpc}$. The second way combines the Planck data
on temperature anisotropies of the cosmic microwave background radiation
\cite{Plan20} with the $\Lambda $CDM model and gives the result $%
H_{0}\approx 67\,\mathrm{km/s/Mpc}$. A number of models have been discussed
in the literature to address this problem, also called Hubble tension (for a
review see \cite{Free21}). In particular, they include the models with
negative cosmological constant (see \cite{Visi19}-\cite{Sen21}). The
maximally symmetric solution of the Einstein field equations with a negative
cosmological constant as the only source of the gravitation is given by
anti-de Sitter (AdS) spacetime. This geometry appears as a ground state in
string theories and in supergravity. It plays an important role in
braneworld models with large extra dimensions and in holographic duality
models relating two theories living in different numbers of spatial
dimensions. An example of the latter is the AdS/CFT correspondence (see, for
example, \cite{Nast15}) establishing the duality between supergravity and
string theories on the AdS bulk and conformal field theory on the AdS
boundary.

Another example for a gravitational source with the negative energy density,
that can play an important role in the expansion of the early universe, is
provided by the vacuum polarization. The vacuum expectation value of the
energy-momentum tensor for quantum fields may break the energy conditions of
the singularity theorems in general relativity (see, e.g., \cite{Birr82}).
This can serve as the key for solving the singularity problems in the
cosmological dynamics. Here we consider the cosmological dynamics for both
cases of positive and negative energy densities. Various combinations of
cosmological constant and of a source with barotropic equation of state will
be studied. Having in mind possible applications in higher-dimensional
models, in particular, motivated by string theories, the discussion is
presented for a general number of spatial dimensions. The qualitative
evolution in cosmological models with scalar fields having negative
potentials has been considered in \cite{Saha97}-\cite{Feld02}. Various cases
of exact solutions to Friedmann equations in general number of spatial
dimensions were discussed in \cite{Chen14} by using Chebyshev's theorem.
Cosmological solutions in (3+1)-dimensional spacetime with a single positive
and negative energy component in a flat universe and for a negative energy
component in a curved universe have been described in \cite{Nemi15}.

The present paper is organized as follows. In the next section we present
the cosmological equations and some qualitative features. The solutions for
flat model with a cosmological constant and barotropic matter are given in
section \ref{sec:Flat}. They serve as past or future attractors for models
with curved space and include various special cases previously considered in
the literature. In section \ref{sec:Curved} we discuss models with curved
space. First, the general solutions are presented in terms of the
hypergeometric function for models with zero cosmological constant. Various
special cases where the time-dependence of the scale factor is expressed in
terms of elementary functions were discussed in the literature. Then we
describe the qualitative evolution in models with curved space driven by a
cosmological constant and barotropic matter source.

\section{Cosmological equations}

\label{sec:Eqs}

We consider $(D+1)$-dimensional background spacetime described by the FRW
line element
\begin{equation}
ds^{2}=N^{2}(t)dt^{2}-a^{2}(t)\left( \frac{dr^{2}}{1-kr^{2}}+r^{2}d\Omega
_{D-1}^{2}\right) ,  \label{ds2}
\end{equation}%
where $d\Omega _{D-1}^{2}$ is the line element on a unit sphere $S^{D-1}$
and $k=0,\pm 1$. The choices $N(t)=1$ and $N(t)=a(t)$ correspond to the
synchronous and conformal time coordinates, respectively. Depending on the
equation of state the first or the second choice of the time coordinate is
convenient to present the cosmological solutions in simpler form. Assuming
that the dynamics is governed by General Relativity in $(D+1)$-dimensional
spacetime, the set of cosmological equations takes the form%
\begin{eqnarray}
\frac{d}{dt}\left( \frac{\dot{a}}{a}\right) +\frac{\dot{a}}{a}\left( D\frac{%
\dot{a}}{a}-\frac{\dot{N}}{N}\right) +(D-1)N^{2}\frac{k}{a^{2}} &=&\frac{%
8\pi G_{D}}{D-1}N^{2}\left( \varepsilon -p\right) ,  \notag \\
\left( \frac{\dot{a}}{a}\right) ^{2}+\frac{N^{2}k}{a^{2}} &=&\frac{16\pi
G_{D}}{D(D-1)}N^{2}\varepsilon ,  \label{Coseq}
\end{eqnarray}%
where the dot stands for the derivative with respect to $t$, $G_{D}$ is the
gravitational constant in $(D+1)$-dimensional spacetime, $\varepsilon $ is
the energy density and $p$ is the pressure for the sources driving the
cosmological evolution. The latter two quantities obey the equation $\dot{%
\varepsilon}+D(\dot{a}/a)(\varepsilon +p)=0$ which is obtained from the
covariant conservation equation for the energy-momentum tensor. This
relation can also be obtained from (\ref{Coseq}). For the second derivative
of the scale factor we get%
\begin{equation}
\frac{\ddot{a}}{a}-\frac{\dot{N}}{N}\frac{\dot{a}}{a}=-\frac{8\pi G_{D}}{D-1}%
N^{2}\left( p+\frac{D-2}{D}\varepsilon \right) .  \label{Accel}
\end{equation}%
From this relation it follows that the accelerated expansion in terms of the
synchronous time coordinate ($N(t)=1$) is obtained under the condition $%
p<(2-D)\varepsilon /D$. The latter condition is satisfied by the positive
cosmological constant $\Lambda $ with the energy density $\varepsilon
_{\Lambda }=\Lambda /(8\pi G_{D})$ and pressure $p_{\Lambda }=-\varepsilon
_{\Lambda }$.

In the discussion below we assume that the matter source contains two parts
with $\varepsilon =\varepsilon _{\Lambda }+\varepsilon _{m}$ and $%
p=p_{\Lambda }+p_{m}$. Here, the part with the equation of state $p_{\Lambda
}=-\varepsilon _{\Lambda }$ corresponds to the cosmological constant $%
\Lambda $ with the constant energy density $\varepsilon _{\Lambda }$ and the
second contribution has an equation of state $p_{m}=w\varepsilon _{m}$ with $%
w=$const. The condition $p<(2-D)\varepsilon /D$ for the second source is
reduced to $w<w_{c}\equiv 2/D-1$ for $\varepsilon _{m}>0$ and to $w>w_{c}$
for $\varepsilon _{m}<0$. From the covariant conservation equation we get%
\begin{equation}
\varepsilon _{m}=\varepsilon _{m0}\left( a/a_{0}\right) ^{-\alpha },
\label{epsm}
\end{equation}%
with the notation%
\begin{equation}
\alpha =D(1+w).  \label{alfa}
\end{equation}%
We will assume that the cosmological constant $\Lambda $ and the constant $%
\varepsilon _{m0}=\varepsilon _{m}|_{a=a_{0}}$ can be either positive or
negative. Note that from the second equation in (\ref{Coseq}) it follows
that one needs to have the condition $16\pi G_{D}\varepsilon \geq
D(D-1)ka^{-2}$ and the total energy density $\varepsilon $ should be
nonnegative in models with $k=0,1$.

Let us consider the qualitative features of the evolution in terms of the
synchronous time. Taking $N(t)=1$, the second equation in (\ref{Coseq}) is
rewritten as%
\begin{equation}
H^{2}+\frac{k}{a^{2}}=\frac{2\Lambda }{D(D-1)}+\frac{16\pi G_{D}\varepsilon
_{m0}}{D(D-1)\left( a/a_{0}\right) ^{\alpha }},  \label{Heq}
\end{equation}%
where $H=\dot{a}/a$ is the Hubble function. From here it follows that for $%
w>-1$ and for a positive cosmological constant the late time evolution
(large values of the scale factor) is dominated by the first term in the
right hand side. In this case the de Sitter solution $a(t)\propto
e^{H_{\Lambda }t}$, with
\begin{equation}
H_{\Lambda }=\sqrt{\frac{2|\Lambda |}{D(D-1)}}  \label{Hlam}
\end{equation}%
(here we consider the case $\Lambda >0$, the notation $H_{\Lambda }$ for $%
\Lambda <0$ is used below), is the future attractor for the general
solution. For a negative cosmological constant, $\Lambda <0$, and for $w>-1$%
, from (\ref{Heq}) we see that with increasing $a$ at some moment $t=t_{%
\mathrm{m}}$ the Hubble function becomes zero. The corresponding value for
the scale factor $a=a_{\mathrm{m}}$ is determined from (\ref{Heq}) putting $%
H=0$. At that moment from the first equation (\ref{Coseq}) we get
\begin{equation}
\dot{H}_{t=t_{\mathrm{m}}}=-Dk\frac{w-w_{c}}{2a_{\mathrm{m}}^{2}}+\frac{1+w}{%
D-1}\Lambda .  \label{Hm}
\end{equation}%
For $k=0,1$ and $w>w_{c}$ the right-hand side is negative and for $t>t_{%
\mathrm{m}}$ one obtains $H<0$ and the initial expansion is followed by the
contraction. The same is the case for $k=-1$ and $-1<w<w_{c}$. For $\alpha
>2 $ and $\varepsilon _{m0}>0$, the early expansion, corresponding to small
values of the scale factor, is dominated by the matter source and the
solutions with flat space serve as attractors for models with $k=\pm 1$.

\section{Cosmological solutions in flat model}

\label{sec:Flat}

\begin{figure}[tbph]
\begin{center}
\epsfig{figure=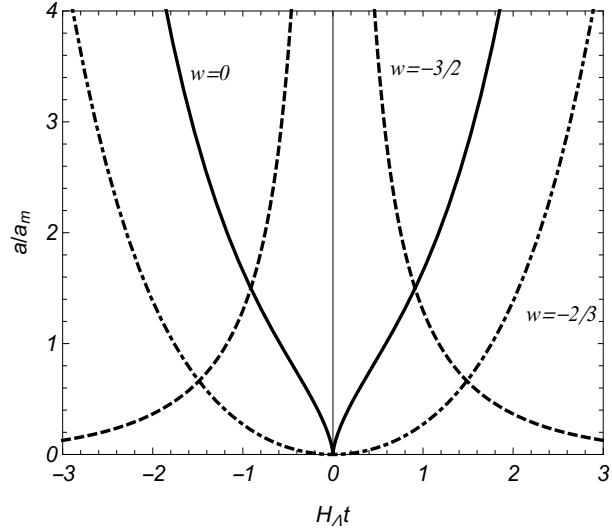,width=8cm,height=7cm}
\end{center}
\caption{The time dependence of the scale factor in the model with $\protect%
\varepsilon _{m0},\protect\varepsilon _{\Lambda }>0$ for $D=3$ and $%
w=0,-2/3,-3/2$. }
\label{fig1}
\end{figure}

Simple exact solutions of the cosmological equations can be found in the
case of flat model, $k=0$. In the absence of the matter source the equation (%
\ref{Heq}) has solutions only for $\Lambda \geq 0$. For positive
cosmological constant the de Sitter solution, $a(t)\propto e^{\pm H_{\Lambda
}t}$, is obtained. To see the influence of the matter source, first we
consider the case of positive cosmological constant and positive matter
density, corresponding to $\varepsilon _{m0},\varepsilon _{\Lambda }>0$. In
the synchronous time coordinate, for the Hubble function we get
\begin{equation}
H=\pm H_{\Lambda }\sqrt{1+\left( \frac{a_{\mathrm{m}}}{a}\right) ^{\alpha }}%
,\;\frac{a_{\mathrm{m}}}{a_{0}}\equiv \left\vert \frac{\varepsilon _{m0}}{%
\varepsilon _{\Lambda }}\right\vert ^{1/\alpha },  \label{Hsol1}
\end{equation}%
with $\alpha $ defined by (\ref{alfa}). The integration of this equation
leads to the following expressions for the Hubble function and the scale
factor:%
\begin{equation}
H(t)=\pm H_{\Lambda }\coth \left( \beta \left\vert t\right\vert \right)
,\;a(t)=a_{\mathrm{m}}\sinh ^{2/\alpha }\left( \beta \left\vert t\right\vert
\right) ,  \label{asol1}
\end{equation}%
where%
\begin{equation}
\beta =\frac{1}{2}|\alpha |H_{\Lambda }=\left\vert 1+w\right\vert \sqrt{%
\frac{D|\Lambda |}{2(D-1)}}.  \label{beta}
\end{equation}%
For $w>-1$ the solution (\ref{asol1}) for the scale factor coincides with
that found in \cite{Chen14}. In that case and for expansion models one has $%
0<t<\infty $ with the upper sign in the expression for the Hubble function.
At late times, $\beta t\gg 1$, one has an approximately de Sitter expansion
with $a(t)\propto e^{H_{\Lambda }t}$. Near the singularity point $t=0$ we
obtain $a(t)\propto |t|^{2/\alpha }$. The case $w<-1$ corresponds to the
phantom phase (for the effective phantom phase generated by different types
of sources see \cite{Eliz05}). In this case $\alpha <0$ and for the
expansion models we have $-\infty <t<0$. The point $t=0$ corresponds to the
Big Rip singularity. The universe starts with de Sitter expansion $%
a(t)\propto e^{H_{\Lambda }t}$, $\beta \left\vert t\right\vert \gg 1$, in
the infinite past and ends the evolution at Big Rip singularity at $t=0$
with the behavior $a(t)\propto |t|^{-2/|\alpha |}$. In figure \ref{fig1} we
have plotted the ratio $a/a_{\mathrm{m}}$ versus $H_{\Lambda }t$ for $D=3$.
The full and dashed curves correspond to the values $w=0$ (dust matter), $%
w=-2/3$ and $w=-3/2$ (phantom matter). Note that under certain conditions
(see \cite{Odin20}) the energy density for the axion field scales as $%
\varepsilon _{\mathrm{axion}}\sim 1/a^{3}$ and the corresponding dynamics is
described by the curve with $w=0$ in figure \ref{fig1} (the cosmological
dynamics with the axion field and holographic dark energy has been recently
discussed in \cite{Saha21a}). For expanding models we have $0<t<+\infty $
for sources with $w>-1$ and $-\infty <t<0$ for $w<-1$. The singular point $%
t=0$ corresponds to the Big Bang in the first case and to the Big Rip in the
second case. For $w>w_{c}$ and $w<-1$ one has $\dot{a}|_{t=0}=\infty $ and
for $-1<w<w_{c}$ we get $\dot{a}|_{t=0}=0$. We see that for $\varepsilon
_{m0},\varepsilon _{\Lambda }>0$ all the flat models contain singularities.

\begin{figure}[tbph]
\begin{center}
\epsfig{figure=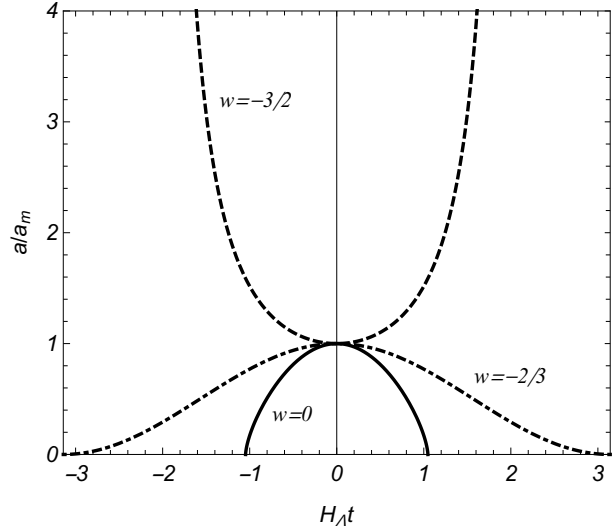,width=8cm,height=7cm}
\end{center}
\caption{The same as in figure \protect\ref{fig1} for the model mith $%
\protect\varepsilon _{\Lambda }<0<\protect\varepsilon _{m0}$. }
\label{fig2}
\end{figure}

Next we consider the case $\varepsilon _{\Lambda }<0<\varepsilon _{m0}$. For
the Hubble function we find%
\begin{equation}
H=\pm H_{\Lambda }\sqrt{\left( a_{\mathrm{m}}/a\right) ^{\alpha }-1}.
\label{Hsol2}
\end{equation}%
The time dependences for the Hubble function and for the scale factor read%
\begin{equation}
H(t)=\pm H_{\Lambda }\tan \left( \beta |t|\right) ,\;a(t)=a_{\mathrm{m}}\cos
^{2/\alpha }\left( \beta |t|\right) ,  \label{asol2}
\end{equation}%
with $-\pi /2\beta <t<\pi /2\beta $. For $w>-1$ this solution coincides with
that presented in \cite{Chen14}. The authors of \cite{Chen14} emphasize that
the solution (\ref{asol2}) gives rise to a periodic universe. However, it
should be noted that, though the function $a(t)$ in (\ref{asol2}) is
periodic with the period $t_{L}=\pi /\beta $, the periods are separated by
singular points $|t|=\pi (l+1/2)/\beta $, $l=0,1,2,\ldots $, and the
evolution pieces separated by those points present the copies of the same
universe with a finite lifetime $t_{L}$ (for discussion of various types of
singularities in the cosmological context see, for example, \cite%
{Noji05,Bamb12,Kame13}). The dependence of the scale factor on the
synchronous time coordinate, described by (\ref{asol2}), is depicted in
figure \ref{fig2} for $D=3$ and $w=0,-2/3,-3/2$. In models with $w>-1$ the
expansion phase with $-\pi /2\beta <t<0$ is followed by the contraction one
for $0<t<\pi /2\beta $. The maximal value of the scale factor is determined
by (\ref{Hsol1}). For sources with $w<-1$ the same relation determines the
minimal value of the scale factor. Similar to the previous case, the flat
models contain singularities for all values of the parameters.

\begin{figure}[tbph]
\begin{center}
\epsfig{figure=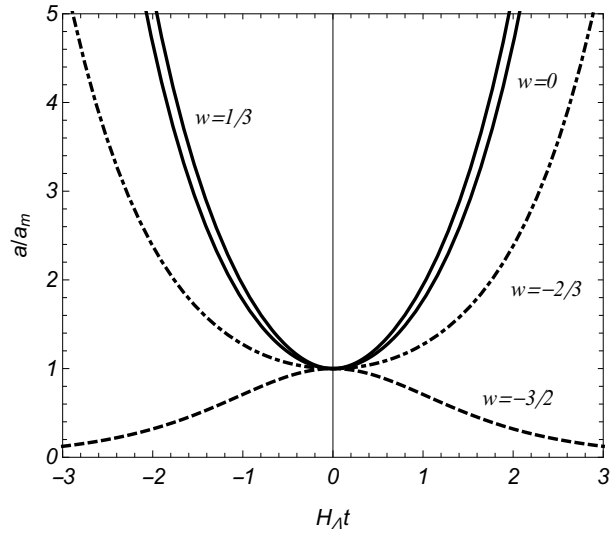,width=8cm,height=7cm}
\end{center}
\caption{The scale factor versus the time coordinate in the model with $%
\protect\varepsilon _{m0}<0<\protect\varepsilon _{\Lambda }$ for $D=3$ and $%
w=1/3,0,-2/3,-3/2$. }
\label{fig3}
\end{figure}

Now we turn to the case $\varepsilon _{m0}<0<\varepsilon _{\Lambda }$. The
Hubble function is expressed as%
\begin{equation}
H=\pm H_{\Lambda }\sqrt{1-\left( a_{\mathrm{m}}/a\right) ^{\alpha }},
\label{Hsol3}
\end{equation}%
where $a_{\mathrm{m}}$ is the minimal (maximal) value of the scale factor
for $w>-1$ ($w<-1$). The time dependence is given by the formulas%
\begin{equation}
H=\pm H_{\Lambda }\tanh \left( \beta \left\vert t\right\vert \right) ,\;a=a_{%
\mathrm{m}}\cosh ^{2/\alpha }\left( \beta \left\vert t\right\vert \right) ,
\label{asol3}
\end{equation}%
with $-\infty <t<+\infty $. The time dependence of the scale factor given by
(\ref{asol3}) is plotted in figure \ref{fig3} for the values of the
parameters $D=3$ and $w=1/3,0,-2/3,-3/2$. The models in this case have no
singularities. The value $a=a_{\mathrm{m}}$ determines the minimum/maximum
value of the scale factor. Note that flat cosmological models with $%
\varepsilon _{m0},\varepsilon _{\Lambda }<0$ are not allowed by the equation
(\ref{Heq}). The corresponding models with curved space will be discussed in
the next section.

In \cite{Saha21,Saha22} it has been shown that in de Sitter spacetime the
vacuum expectation value of the energy-momentum tensor for a conformally
coupled massless scalar field in the hyperbolic vacuum has the form
\begin{equation}
\left\langle T_{i}^{k}\right\rangle =\varepsilon _{\Lambda }\mathrm{diag}%
\left( 1,1,\cdots ,1\right) +\frac{C_{D}}{a^{D+1}}\mathrm{diag}\left( 1,-%
\frac{1}{D},\cdots ,-\frac{1}{D}\right) ,  \label{TikVac}
\end{equation}%
where the first term in the right-hand side corresponds to a cosmological
constant and the constant $C_{D}$ is negative. The second term can be
identified with the source we have considered above having the equation of
state $p_{m}=$ $\varepsilon _{m}/D$. Hence, for this source one gets $w=1/D$
and $\varepsilon _{m0}<0$. The corresponding cosmological solution is given
by (\ref{asol3}) with $\alpha =D+1$ and $\beta =\left( D+1\right) H_{\Lambda
}/2$. The time dependence of the scale factor for $D=3$ is presented in
figure \ref{fig3} by the curve with $w=1/3$. The corresponding models are
nonsingular.

\section{Cosmological dynamics in models with curved space}

\label{sec:Curved}

Passing to the models with curved space, first let us recall the well-known
solutions in the absence of matter sources. The models with $k=1$ are
allowed only in the case $\Lambda >0$ and the corresponding solution for the
scale factor is given by $a=\cosh \left( H_{\Lambda }t\right) /H_{\Lambda }$%
. For $k=-1$ and $\Lambda >0$ the solution reads $a=\sinh \left( H_{\Lambda
}|t|\right) /H_{\Lambda }$. For $k=-1$ and negative cosmological constant we
have the solution $a=\sin \left( H_{\Lambda }|t|\right) /H_{\Lambda }$. Note
that in models with $k=-1$ and $H_{\Lambda }|t|\ll 1$ the evolution is
approximated by linear scale factor $a(t)=|t|$. The latter describes a flat
spacetime and corresponds to the Milne universe.

Another special case corresponds to the absence of cosmological constant.
From the equation (\ref{Heq}) we get%
\begin{equation}
\frac{dy}{d(t/a_{0})}=\pm \sqrt{\gamma _{0}y^{\gamma }-k},  \label{Heq0}
\end{equation}%
with the notations%
\begin{equation}
y=\frac{a}{a_{0}},\;\gamma _{0}=\frac{16\pi G_{D}\varepsilon _{m0}a_{0}^{2}}{%
D(D-1)},\;\gamma =D(w_{c}-w).  \label{gam}
\end{equation}%
Separating the variables, the integrals in (\ref{Heq0}) can be expressed in
terms of the incomplete beta function $B_{z}(u,v)$. Presenting the latter
through the hypergeometric function $F(a,b;c;z)$ (see, for example, \cite%
{Olve10}), for the models with $k=-1$ we find%
\begin{equation}
t=\frac{a}{\sqrt{\gamma _{0}y^{\gamma }+1}}F\left( \frac{1}{2},1;1+\frac{1}{%
\gamma },\frac{\gamma _{0}y^{\gamma }}{\gamma _{0}y^{\gamma }+1}\right) .
\label{tkm1}
\end{equation}%
In a similar way, for the models with $k=1$ the integration gives%
\begin{equation}
t=\frac{2a_{0}}{\gamma \gamma _{0}^{1/\gamma }}\left( 1-\frac{1}{\gamma
_{0}y^{\gamma }}\right) ^{1/2}F\left( \frac{1}{2},\frac{1}{2}+\frac{1}{%
\gamma };\frac{3}{2};1-\frac{1}{\gamma _{0}y^{\gamma }}\right) .
\label{tkp1}
\end{equation}%
The various special cases of these general formulas have been considered in
the literature. In particular, the examples when for general number of
spatial dimension the solutions are expressed in terms of elementary
functions have been discussed in \cite{Chen14}.

Now we turn to the general case of models with curved space in the presence
of a cosmological constant and barotropic matter. The equation (\ref{Heq})
is rewritten as%
\begin{equation}
\frac{dx}{dt}=\pm H_{\Lambda }\sqrt{s_{\Lambda }x^{2}+bx^{\gamma }-k},
\label{Heq1}
\end{equation}%
where
\begin{equation}
x=H_{\Lambda }a,s_{\Lambda }=\mathrm{sgn}(\Lambda ),  \label{slam}
\end{equation}%
and%
\begin{equation}
b=\frac{16\pi G_{D}\varepsilon _{m0}a_{0}^{2}}{D(D-1)}\left( H_{\Lambda
}a_{0}\right) ^{\gamma }.  \label{b}
\end{equation}%
Simple solutions are found for the special case of the source with $w=w_{c}$%
. For $\Lambda >0$ and $\gamma _{0}-k>0$ the solution has the form $x=\sqrt{%
\gamma _{0}-k}\sinh \left( H_{\Lambda }|t|\right) $. In the case $\Lambda >0$
and $\gamma _{0}-k<0$, the solution reads $x=\sqrt{k-\gamma _{0}}\cosh
\left( H_{\Lambda }t\right) $. For $\Lambda <0$ one needs to have $\gamma
_{0}-k>0$ and the corresponding solution is given by $x=\sqrt{\gamma _{0}-k}%
\sin \left( H_{\Lambda }t\right) $, $0<t<\pi /H_{\Lambda }$. For $\gamma
_{0}=0$ the first two solutions are reduced to the de Sitter solutions.

We will denote by $x=x_{\mathrm{m}}>0$ the value of the function $x(t)$ at
its possible extremum, $dx/dt|_{x=x_{\mathrm{m}}}=0$. The extrema are zeros
of the expression under the square root in (\ref{Heq1}). Taking the
corresponding value of the time coordinate as $t=0$ and expanding near the
extremum we get%
\begin{equation}
\frac{a(t)}{a_{\mathrm{m}}}\approx 1+\frac{D}{4}\left[ s_{\Lambda }\left(
1+w\right) +\frac{w_{c}-w}{x_{\mathrm{m}}^{2}}k\right] \left( H_{\Lambda
}t\right) ^{2},  \label{aext}
\end{equation}%
where $a_{\mathrm{m}}=x_{\mathrm{m}}/H_{\Lambda }$. The nature of the
extremum (minimum or maximum) is determined by the sign of the expression in
the square brackets. Note that for the extremum we have $bx_{\mathrm{m}%
}^{-\alpha }=k/x_{\mathrm{m}}^{2}-s_{\Lambda }$. In the definition of the
constant $b$ we have taken $a_{0}=a(t_{0})$ and $\varepsilon
_{m0}=\varepsilon _{m}(t_{0})$ for a fixed time $t=t_{0}$. Taking $t_{0}=t_{%
\mathrm{m}}$, where $t_{\mathrm{m}}$ corresponds to the extremal value $x_{%
\mathrm{m}}$, $x(t_{\mathrm{m}})=x_{\mathrm{m}}$, from (\ref{Heq}) we get
the following relation%
\begin{equation}
x_{\mathrm{m}}^{2}=\frac{s_{\Lambda }k}{1+\varepsilon _{\mathrm{(m)}%
}/\varepsilon _{\Lambda }},  \label{xm}
\end{equation}%
where $\varepsilon _{\mathrm{(m)}}=\varepsilon (t_{\mathrm{m}})$ is the
matter energy density at the extremum point. Note that, assuming the
presence of the extremum $x=x_{\mathrm{m}}$, the equation (\ref{Heq1}) is
written as%
\begin{equation}
\frac{dy}{dt}=\pm H_{\Lambda }\sqrt{s_{\Lambda }\left( y^{2}-1\right) +\frac{%
\varepsilon _{\mathrm{(m)}}}{|\varepsilon _{\Lambda }|}\left( y^{\gamma
}-1\right) },  \label{Heq2}
\end{equation}%
with $y=x/x_{\mathrm{m}}=a/a_{\mathrm{m}}$.

Let us consider different combinations of the signs for the energy
densities. For $\varepsilon _{m0},\varepsilon _{\Lambda }>0$ and $w>w_{c}$,
the early dynamics, corresponding to small values of $x$, is dominated by
the source with the energy density $\varepsilon _{m}$ and the expansion law
is close to the one for the flat model. At late times, corresponding to $%
x\gg 1$, the expansion is dominated by the cosmological constant and, again,
the curvature term is subdominant. The solution corresponding to the flat
model is the future attractor for models with curved space. The dependence
of the scale factor on time coordinate is qualitatively similar to that
depicted in figure \ref{fig1} for $w=0$.

For $\varepsilon _{m0},\varepsilon _{\Lambda }>0$, $w<w_{c}$, and $k=-1$,
the early dynamics ($x\ll 1$) for expanding models is dominated by the
curvature term and $a(t)\approx t$, $t\rightarrow 0$. As it has been
mentioned above, the spacetime with $k=-1$ and $a(t)=t$ is flat and
corresponds to the Milne universe. The matter energy density behaves as $%
\varepsilon _{m}\propto x^{-\alpha }$ and for $w>-1$ it diverges at $t=0$
like $\varepsilon _{m}\sim t^{-\alpha }$. In the model with $\varepsilon
_{m0},\varepsilon _{\Lambda }>0$, $w<w_{c}$, and $k=1$ the scale factor has
a minimal value that corresponds to the zero $x=x_{\mathrm{m}}$ of the
expression in the right-hand side of (\ref{Heq1}). At this point the Hubble
function becomes zero. The time-dependence of the scale factor near the
minimum, $H_{\Lambda }t\ll 1$, is given by (\ref{aext}) with $s_{\Lambda }=1$
and $a_{\mathrm{m}}=a_{\mathrm{min}}$. At late times of the expansion, $x\gg
1$, the curvature term in (\ref{Heq1}) can be ignored and the cosmological
dynamics is well approximated by the solutions for flat model (see the
graphs with $w=-2/3,-3/2$ in figure \ref{fig1}). We conclude that the models
with $\varepsilon _{m0},\varepsilon _{\Lambda }>0$, $w<w_{c}$ and $k=1$ are
nonsingular.

Let us turn to the models with $\varepsilon _{\Lambda }<0<\varepsilon _{m0}$%
. For $w>w_{c}$ the maximum allowed value for $x$ is determined by the zero $%
x=x_{\mathrm{m}}$ of the right-hand side in (\ref{Heq1}). The asymptotic
behavior near the maximum is described by (\ref{aext}) with $s_{\Lambda }=1$
and $a_{\mathrm{m}}=a_{\mathrm{max}}$. For $x\ll 1$, in the right-hand side
of (\ref{Heq1}) we can omit the curvature term and $x^{2}$. The scale factor
is approximated by the solution for the flat model and near the Big Bang,
corresponding to $t=-t_{1}$, $t_{1}>0$, one has $a(t)\propto \left(
t+t_{1}\right) ^{2/\alpha }$. The model has finite lifetime $2t_{1}$ and the
corresponding time-dependence of the scale factor is qualitatively similar
to that for the flat model presented by the graph with $w=0$ in figure \ref%
{fig2}.

For $\varepsilon _{\Lambda }<0<\varepsilon _{m0}$ and $-1<w<w_{c}$ the
function $x(t)$ has a maximal allowed value $x=x_{\mathrm{m}}$ determined by
the zero of the right-hand side in (\ref{Heq1}). Taking $x(0)=x_{\mathrm{m}}$%
, near the maximum point we have the approximation (\ref{aext}) with $%
s_{\Lambda }=1$ and $a_{\mathrm{m}}=a_{\mathrm{max}}$. For $k=-1$, the
models start the expansion at $t=-t_{1}$ with the scale factor $a(t)\approx
t+t_{1}$ and the behavior of the scale factor is close to the one for the
Milne universe. The expansion is stopped at $t=0$ and for $t>0$ the model
enters the contraction phase. The latter is ended at $t=t_{1}$ with $%
a(t)\approx t_{1}-t$. Hence, the $k=-1$ models have lifetime $2t_{1}$ and
the Milne universe is the past and future attractor for the corresponding
dynamics. Note that, though the first derivative of the scale factor is
finite at the points $t=\pm t_{1}$ ($|\dot{a}|_{t=\pm t_{1}}=1$), the matter
energy density diverges at those points as $\varepsilon _{m}\sim 1/|t\pm
t_{1}|^{D(1+w)}$. The models with $k=1$ start the expansion from the finite
value of the scale factor $a_{\mathrm{min}}$ at $t=-t_{\mathrm{min}}$. At
that point $\dot{a}(-t_{\mathrm{min}})=0$. At $t=0$ the scale factor takes
its maximal value $a_{\mathrm{max}}=x_{\mathrm{m}}/H_{\Lambda }$ and then it
enters into the contraction phase. Near the maximum we have the
approximation (\ref{aext}). The evolution is ended at $t=t_{\mathrm{min}}$
with $a=a_{\mathrm{min}}$ and $\dot{a}(t_{\mathrm{min}})=0$. Hence, in this
case we have nonsingular evolution for $-t_{\mathrm{min}}\leq t\leq t_{%
\mathrm{min}}$. Joining the evolutionary pieces with duration $2t_{\mathrm{%
min}}$, we obtain a model with periodically oscillating scale factor in the
limits $a_{\mathrm{min}}\leq a\leq a_{\mathrm{max}}$ for $-\infty <t<+\infty
$.

In models with $\varepsilon _{\Lambda }<0<\varepsilon _{m0}$, $w<-1$, and
for large values of $x$ the expansion law is close to the one for the flat
model and the corresponding behavior is qualitatively close to the one given
by the curve with $w=-3/2$ in figure \ref{fig2}. For small values of $x$ and
for models with $k=-1$ the expansion/contraction law is approximated by $%
a(t)\approx |t|$. At $t=0$ the matter energy density vanishes as $%
\varepsilon _{m}\sim |t|^{D|1+w|}$. In models with $k=1$ the scale factor
has a minimum value $a=a_{\mathrm{min}}$ determined by the zero of the
right-hand side in (\ref{Heq1}) and the evolution for all values of $x\geq
x_{\mathrm{min}}$ is qualitatively similar to that described by the curve
with $w=-3/2$ in figure \ref{fig2}. The expansion models have Big Rip
singularity.

Now let us consider models with the energy densities in the range $%
\varepsilon _{m0}<0<\varepsilon _{\Lambda }$.\ For $w>w_{c}$ the scale
factor has a minimal value $a=a_{\mathrm{min}}$ which is determined by the
zero of the right-hand side in (\ref{Heq1}). Taking $t=0$ for the
corresponding value of the time coordinate, near the minimum one has the
approximation (\ref{aext}) with $a_{\mathrm{m}}=a_{\mathrm{min}}$ and $x_{%
\mathrm{m}}=x_{\mathrm{min}}$. For $w>w_{c}$ and for large values of $x$ the
evolution is approximated by de Sitter spacetime with the Hubble constant $%
H_{\Lambda }$. The behavior of the scale factor is qualitatively similar to
that depicted in figure \ref{fig3} by the curves with $w=0,1/3$ and the
corresponding models have no singularities. An example with positive
cosmological constant, negative matter energy density and the equation of
state parameter $w=1/D>w_{c}$ is provided by (\ref{TikVac}). In the range $%
-1<w<w_{c}$ and for large values of $x$ the evolution is again dominated by
the cosmological constant with de Sitter spacetime being the past or future
attractor. In the same range for $w$ and for $k=-1$ one gets the approximate
solution $a(t)\approx |t|$ for $H_{\Lambda }|t|\ll 1$, corresponding to the
Milne universe. The matter energy density diverges at $t=0$. For $-1<w<w_{c}$
and $k=1$ the scale factor has a minimal value determined by the zero of the
right-hand side of (\ref{Heq1}). Near that minimum the scale factor is
approximated by (\ref{aext}) and the model is nonsingular. In the range $%
w<-1 $ the scale factor has the maximal value $a_{\mathrm{max}}$ which is
given by the zero of the right-hand side in (\ref{Heq1}). For models with $%
k=-1$ the expansion starts at $t=-t_{1}$ with the asymptotic $a(t)\approx
t+t_{1}$ (curvature dominated expansion) and ends at $t=0$ with the
asymptotic given by (\ref{aext}). The expansion phase is followed by the
contraction for $0<t<t_{1}$ with $a(t)\approx t_{1}-t$ near $t=t_{1}$. For
models with $k=1$ the scale factor varies between two nonzero values $0<a_{%
\mathrm{min}}\leq a\leq a_{\mathrm{max}}<\infty $. The corresponding models
are nonsingular and can be extended for $t\in (-\infty ,+\infty )$. The
qualitative dynamics is similar to that we have described above for the case
$k=1$, $\varepsilon _{\Lambda }<0<\varepsilon _{m0}$, $-1<w<w_{c}$.

Finally, for $\varepsilon _{m0},\varepsilon _{\Lambda }<0$, in accordance
with (\ref{Heq}), the models with $k=0$ and $k=1$ are not allowed. Let us
consider the features of the cosmological dynamics in this case for $k=-1$.
For $w>w_{c}$, from the condition for the positivity of the expression under
the square root in (\ref{Heq1}), we can see that the model is allowed under
the constraint
\begin{equation}
|b|<\frac{2}{D\left( w-w_{c}\right) }\left( \frac{w-w_{c}}{w+1}\right)
^{\alpha /2}.  \label{Cond1}
\end{equation}%
This condition restricts the allowed values for the negative energy density $%
\varepsilon _{m0}$. In the range determined by (\ref{Cond1}), the right-hand
side of (\ref{Heq1}) has two zeros and they determine the minimal and
maximal values for the scale factor, $0<a_{\mathrm{min}}\leq a(t)\leq a_{%
\mathrm{max}}$. At those points $\dot{a}=0$ and $H=0$. Near the extrema the
scale factor is approximated by (\ref{aext}) with $s_{\Lambda }=-1$ and $%
k=-1 $. From (\ref{aext}) it follows that%
\begin{equation}
a_{\mathrm{min}}<\frac{1}{H_{\Lambda }}\sqrt{\frac{w-w_{c}}{1+w}}<a_{\mathrm{%
max}}.  \label{alim}
\end{equation}%
For $\varepsilon _{m0},\varepsilon _{\Lambda }<0$ and $w<w_{c}$ the right
hand side of (\ref{Heq1}) has a single zero that determines the maximal
value of the scale factor $a_{\mathrm{max}}=a(0)$. Near the maximum the
scale factor behaves like (\ref{aext}) with $s_{\Lambda }=-1$ and $k=-1$.
For small values of $x$ the dynamics is dominated by the curvature term with
the Milne universe as the asymptotic. The expansion starts at $t=-t_{1}$
with $a(t)\approx t+t_{1}$ and stops at $t=0$ with the maximal value of the
scale factor. The evolution for $0<t<t_{1}$ corresponds to the contraction
phase with the future attractor $a(t)\approx t_{1}-t$. At the points $t=\pm
t_{1}$ the matter energy density vanishes for $w<-1$ and diverges for $%
-1<w<w_{c}$.

\section{Conclusion}

\label{sec:Conc}

We have considered the dynamics of $(D+1)$-dimensional FRW cosmological
models driven by the cosmological constant and the matter source with
barotropic equation of state assuming that the energy densities for those
sources can be either positive or negative. Exact solutions are provided for
models with flat space which include various special cases previously
considered in the literature. In particular, it has been demonstrated that
nonsingular solutions are obtained only for negative energy density of the
matter, regardless the sign of the cosmological constant. The corresponding
scale factor is given by (\ref{asol3}). Another classes of exact solutions,
expressed in terms of the hypergeometric function (see (\ref{tkm1}) and (\ref%
{tkp1})), are obtained for models with curved space in the absence of
cosmological constant. A number of special cases of those solutions, when
they are expressed in terms of elementary function, have been discussed in
the literature (see, for example, \cite{Chen14}). The qualitative evaluation
for models with curved spaces and with a cosmological constant and matter
source has been described in the second part of section \ref{sec:Curved} for
all the values of the equation of state parameter $w$ and for all
combinations of the signs of the energy densities. Depending on the values
of $w$ one can have Big Bang or Big Rip type singularities. We have also
specified nonsingular models with curved space. For $k=1$, nonsingular
modelas are obtained for the following combinations of conditions: (i) $%
(\varepsilon _{0m}>0,\varepsilon _{\Lambda }>0,w<w_{c})$, (ii) $(\varepsilon
_{0m}>0,\varepsilon _{\Lambda }<0,-1<w<w_{c})$, (iii) $(\varepsilon
_{0m}<0,\varepsilon _{\Lambda }>0)$. In models (ii\ ) and $(\varepsilon
_{0m}<0,\varepsilon _{\Lambda }>0,w>-1)$ the evolution of the scale factor,
as a function of time coordinate, is periodically oscillatory in the limits $%
a_{\mathrm{min}}\leq a(t)\leq a_{\mathrm{max}}$. In the remaining cases, the
qualitative evolution of $k=1$ nonsingular models is similar to that
depicted in figure \ref{fig3} for $w=-2/3,0,1/3$. For models with negative
curvature space there exists at least one point on the time axis where the
scale factor becomes zero. Near those points the evolution is dominated by
the matter source for $w>w_{c}$ and by the curvature term for $w<w_{c}$. In
the second case the scale factor is approximated by a linear
expansion/contraction as a function of the time coordinate. At the point
with zero scale factor the matter energy density diverges for $-1<w<w_{c}$
and vanishes for $w<-1$.

We have seen that the negative energy densities for both the cosmological
constant and matter source enlarge the possible scenarios of cosmological
dynamics. Bearing in mind applications in higher-dimensional models, it
would be interesting to generalize the corresponding results for models with
extra compact dimensions. The compactification leads to additional
contributions to the vacuum expectation value of the energy-momentum tensor.
In general, the effective pressures along compact dimensions differ and for
massless conformally coupled fields the topological contributions are
equivalent to barotropic perfect fluid with anisotropic pressures. In
particular, the coefficients $w$ in the respective equations of state may
have different signs. In the corresponding anisotropic cosmological models
one can have an expansion for a part of dimensions and a contraction for the
remaining ones. The analysis of different cosmological scenarios can be done
in a way similar to that we have described above. We can also use the
methods of qualitative analysis of dynamical systems to classify
qualitatively different cosmological models. The corresponding results for a
toroidal compactification will be presented elsewhere.

\section*{Acknowledgments}

A.A.S., R.M.A.,T.A.P. were supported by Grants No. 20RF-059, No. 21AG-1C047
and No. 20AA-1C005 of the Science Committee of the Ministry of Education,
Science, Culture and Sport RA. E.R.B.d.M. is partially supported by CNPQ
under Grant No. 301.783/2019-3.

\end{document}